\documentclass[manuscript]{acmart}
\usepackage{subcaption}
\newcommand\tool{Model LineUpper}
\newcommand\company{IBM}
\AtBeginDocument{%
  \providecommand\BibTeX{{%
    \normalfont B\kern-0.5em{\scshape i\kern-0.25em b}\kern-0.8em\TeX}}}

\copyrightyear{2021} 
\acmYear{2021} 
\setcopyright{acmlicensed}\acmConference[IUI '21]{26th International Conference on Intelligent User Interfaces}{April 14--17, 2021}{College Station, TX, USA}
\acmBooktitle{26th International Conference on Intelligent User Interfaces (IUI '21), April 14--17, 2021, College Station, TX, USA}
\acmPrice{15.00}
\acmDOI{10.1145/3397481.3450658}
\acmISBN{978-1-4503-8017-1/21/04}

\begin{document}

\title{Model LineUpper: Supporting Interactive Model Comparison at Multiple Levels for AutoML}


\author{Shweta Narkar}
\email{narkas@rpi.edu}
\affiliation{%
  \institution{Rensselaer Polytechnic Institute}
  \streetaddress{110 8th Street}
  \city{Troy}
  \country{USA}}
\author{Yunfeng Zhang}
\email{zhangyun@ibm.com}
\affiliation{%
  \institution{IBM Research AI}
  \streetaddress{1101 Kitchawan Road}
  \city{Yorktown Heights}
  \country{USA}}
\author{Q. Vera Liao}
\email{vera.liao@ibm.com}
\affiliation{%
  \institution{IBM Research AI}
  \streetaddress{1101 Kitchawan Road}
  \city{Yorktown Heights}
  \country{USA}}
\author{Dakuo Wang}
\email{dakuo.wang@ibm.com}
\affiliation{%
  \institution{IBM Research AI}
  \streetaddress{1101 Kitchawan Road}
  \city{Yorktown Heights}
  \country{USA}}
\author{Justin D Weisz}
\email{jweisz@us.ibm.com}
\affiliation{%
  \institution{IBM Research AI}
  \streetaddress{1101 Kitchawan Road}
  \city{Yorktown Heights}
  \country{USA}}
  
\makeatletter
\let\@authorsaddresses\@empty
\makeatother


\begin{abstract}
Automated Machine Learning (AutoML) is a rapidly growing set of technologies that automate the model development pipeline by searching model space and generating candidate models. A critical, final step of AutoML is human selection of a final model from dozens of candidates. In current AutoML systems, selection is supported only by performance metrics. Prior work has shown that in practice, people evaluate ML models based on additional criteria, such as the way a model makes predictions. Comparison may happen at multiple levels, from types of errors, to feature importance, to how the model makes predictions of specific instances. We developed \tool{} to support interactive model comparison for AutoML by integrating multiple Explainable AI (XAI) and visualization techniques. We conducted a user study in which we both evaluated the system and used it as a technology probe to understand how users perform model comparison in an AutoML system. We discuss design implications for utilizing XAI techniques for model comparison and supporting the unique needs of data scientists in comparing AutoML models.
\end{abstract}




\maketitle

\section{Introduction}
Although Machine Learning (ML) technologies have permeated numerous domains, development cost and expertise barriers for building ML models remain high~\cite{gartner2020magic,mao2019}. Automated Machine Learning (AutoML) technologies have reduced development costs by generating optimized ML models using novel model selection, feature engineering, and hyperparameter optimization algorithms~\cite{zoller2019survey,liu2020admm,autods}. Recently, AutoML technologies have matured from development efforts at technology companies~\cite{web:googleautoml,web:h2o,web:ibmautoai}. A research field is also emerging on the study and development of tools that support user interactions with AutoML systems~\cite{wang_atmseer_2019,wang2019humanai,autods}. This work demonstrates the importance of retaining human agency in AutoML workflows~\cite{lee2019human, wang_atmseer_2019,wang2020autoai,automationsurvey}. Users desire transparency features to understand how AutoML works~\cite{karl2020,JaimieDrozdal2020}, as well as control features to make adjustments based on their prior knowledge~\cite{wang_atmseer_2019}. 

Existing tools for AutoML focus on \textit{procedural transparency} ~\cite{wang_atmseer_2019,karl2020,JaimieDrozdal2020}, showing the process by which AutoML searches through algorithmic and optimization spaces. This approach aims to provide assurance that the search was thorough, and allows to adjust the search space configuration. But, it remains questionable whether AutoML users, especially less experienced data scientists (like data workers in~\cite{hou2017hacking}), could act on adjusting the AutoML process. Few systems provide \textit{algorithmic transparency}, showing how AutoML-generated models work, such as how they weigh features and judge specific instances. Most AutoML systems use \emph{performance metrics} as basis for model evaluation and ranking. However, research on model analytics and comparison of manually-built models has shown that none of the stakeholders are satisfied to only see performance metrics~\cite{muller2019datascience,sun_dfseer_2020,zhang_manifold_2019,autods,dscommunicate}. They are interested in details such as types of errors, how models perform on specific instances, and a detailed reasoning by which models make predictions.

These observations have motivated recent work to leverage visualization techniques from Explainable AI (XAI) to support model analytics and debugging~\cite{hohman_gamut_2019,kahng2017cti,brooks_featureinsight_2015,strobelt_lstmvis_2018,krause2016interacting}. XAI techniques allow us to understand the inner-workings of an ML model. But, the amount of human effort required to scrutinize an ML model is high, and is exacerbated in the AutoML context in which dozens of models may be produced from a single experiment. Also, it is unclear how users of AutoML conduct model comparison, since some candidate models tend to be similar variants of one another with differences in optimization choices.

We introduce \tool{}, a visualization tool that provides transparency into candidate models generated by AutoML. \tool{} allows users to interactively compare AutoML models based on multiple aspects of their function and behavior. In a user study with 14 data scientists, we learned that \tool{} helped participants select models based on different criteria such as types of errors and alignment with domain knowledge. Our work highlights the need for algorithmic transparency, evaluates how XAI techniques can support this need, and sheds light on the unique design requirements of AutoML systems.









\section{\tool{}}
The design of \tool{} was informed by prior work on model comparison outside of AutoML context \cite{sun_dfseer_2020,wang_atmseer_2019,zhang_manifold_2019}, as well as many discussions with expert AutoML users. Current AutoML systems~\cite{web:ibmautoai,web:h2o} provide users with overall model performance metrics, but treat individual models as opaque boxes. Our design goal is to enable users to open the opaque boxes and engage in model comparison based on: 1) selected instances of interest or subsets of data; and 2) explanations of how models make predictions. Recent work has introduced instance-level investigation \cite{alsallakh_visual_2014-1,amershi_modeltracker_2015-1,choo_ivisclassifier_2010,ren_squares_2017} and XAI techniques \cite{hohman_gamut_2019,brooks_featureinsight_2015,google_pair_what-if_2018} for model analytics tools, demonstrating their effectiveness in supporting debugging tasks for a single model, and engendering user trust and confidence in the final outcome. But, with one recent exception~\cite{zhang_manifold_2019}, these two capabilities have not been utilized for model comparison. 

\begin{figure*}
    \begin{minipage}{0.45\textwidth}
        \begin{subfigure}{\linewidth}
            \includegraphics[width=\textwidth]{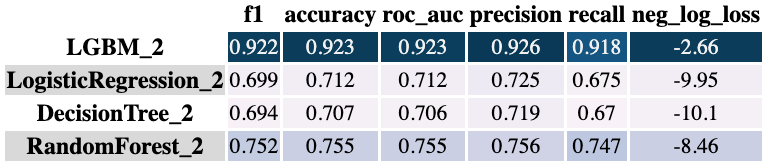}
            \caption{Screenshot of the Metrics Table showing metrics for four selected models.}
            \label{fig:metrics-table}
        \end{subfigure}
        \par\bigskip
        \begin{subfigure}{\linewidth}
            \includegraphics[width=\textwidth]{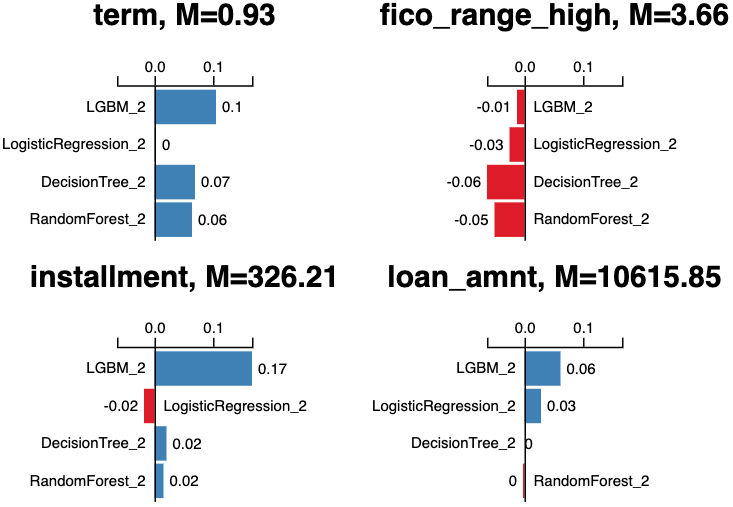}
            \caption{Partial screenshot of the Feature Importance Comparison View showing 4 of 21 FI plots.}
            \label{fig:lfc}
        \end{subfigure}
    \end{minipage}
    \hfill
    \begin{minipage}{0.5\textwidth}
    \begin{subfigure}{\linewidth}
        \includegraphics[width=\textwidth]{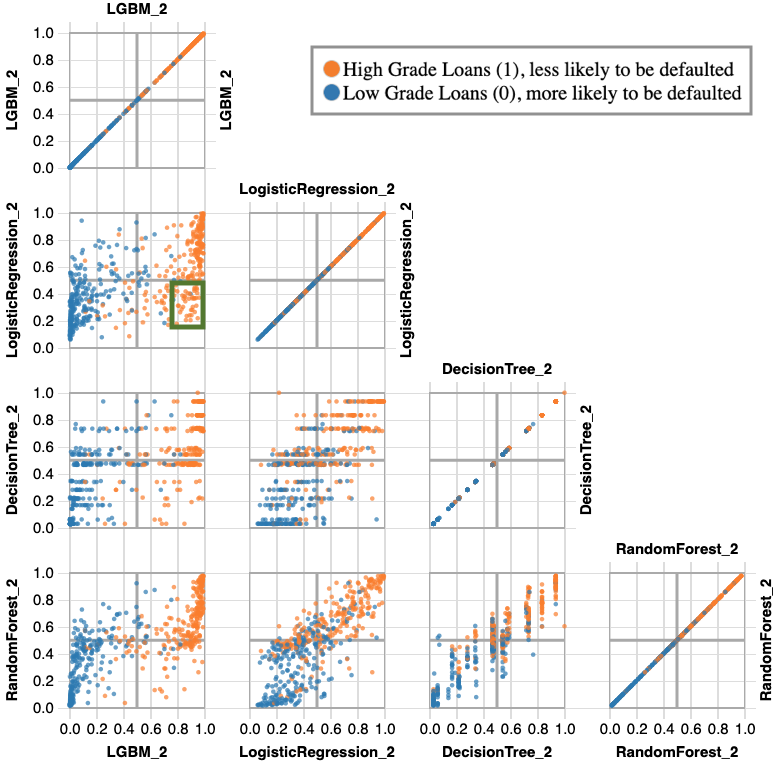}
        \caption{Screenshot of the Probability Scatterplot Matrix displaying pairwise comparisons of 4 models.}
        \label{fig:psm}
    \end{subfigure}
    \end{minipage}
    \caption{The three primary components of \tool{}.}
    \label{fig:toolUI}
\end{figure*}

To support explainability and instance-level investigation, \tool{} consists of three views, seen in Figure~\ref{fig:toolUI}: (a) metrics table, (b) feature importance comparison view, and (c) probability scatterplot matrix. In addition, there are legends for the plots and a control panel that allows users to select models to be displayed.

We use a loan risk modeling task to illustrate the functions of \tool{} and conduct the user study, while in practice the system works with other data and tasks. The models were generated by \company's AutoML system, and were trained with a subset of data published by the LendingClub Corporation~\cite{lendingclub_lendingclub_2020}. Our data set has 47 features and 11,553 instances, each representing a loan application. The model's task is to predict the grade of application (likelihood of repayment). We created a balanced binary label that indicates whether a loan was high grade (Grades A and B in the original dataset) or low grade (Grades C to G). As AutoML iteratively applies variations of optimization to different ML algorithms, \tool{} indicates the algorithm and optimization variants in the model name.  For example, ``LGBM\_2: Hyperparameter Optimization'' indicates a light gradient boosted model (LGBM) with type 2 optimization (hyperparameter optimization).


\subsubsection{Metrics Table}
The metrics table (Figure~\ref{fig:metrics-table}) supports comparing models by overall performance metrics. The set of metrics vary based on type of prediction task (e.g., classification vs. regression). For binary classification tasks, \tool{}  computes common metrics such as F1, accuracy, ROC AUC, etc. Each row corresponds to one model that user has selected to compare. The cells are shaded based on the ranking of values within a column, which guides the comparison and suggests cells for further examination.

 \subsubsection{Feature Importance Comparison View}

Feature importance (FI) is a popular XAI technique that explains a model by how much impact each feature has on its predictions~\cite{strumbelj_explaining_2014,ribeiro_why_2016}. FI can act as both global and local explanations: global FI shows how the model weighs different features in general, whereas local FI explains a model's prediction for a specific instance based on how the model weighed that instance's features. The FI comparison view (Figure~\ref{fig:lfc}) can shift between showing global FI, when no instance is selected, and local FI, when user selects data points in the probability scatterplot matrix  (Section~\ref{sec:psm}). Global FI values are obtained via SciKit-Learn's~\cite{scikit-learn} \textit{feature\_importances\_} when available, or in the case of regression models, taking the absolute value of feature weights. These values are then normalized to allow comparison across different algorithms. Local FI values are generated using the SHAP Python library~\cite{lundberg_unified_2017}, which produces sensitivity-analysis based explanations~\cite{strumbelj_explaining_2014}.

To support comparing FI across models, we visualize FI values of all models for a single feature in one panel and sort the panels by average global FI. When a group of points are selected, we plot average local FI value of them for each feature. We also show average feature \textit{value} of the selected points in the title. In Figure~\ref{fig:lfc}, the selected points have an average installment of \$326.21. The LGBM model considers their installment scores to be a positive indicator of a high-grade loan, whereas other models consider the scores to be somewhat neutral.

 \subsubsection{Probability Scatterplot Matrix}
 \label{sec:psm}

The matrix is inspired by Manifold~\cite{zhang_manifold_2019}, which supports comparison of model pairs and identifies instances of potential interest. In Figure~\ref{fig:psm}, each single panel shows predicted probabilities of instances in the test data set for target class (high grade loans) by a model pair. Each dot represents a data instance, and its $(x, y)$ coordinates correspond to probability that the instance is predicted to be high grade by the two models. The quadrants and color coding present the true/false positives/negatives of the two models' predictions. They show instances on which the two models agree (quadrants 1 and 3) or disagree (quadrants 2 and 4). A blue dot in quadrant 1 indicates a false positive error for both models, whereas a blue dot in quadrant 2 indicates a false positive error for just the model on the $y$ axis. The distribution pattern of the dots helps to compare the confidence of two models, as denser lines at the ends of an axis indicate a more confident model corresponding to the axis (LGBM model in Figure~\ref{fig:psm}).

The matrix serves as an entry point for instance-level investigation, which could be useful for users as they might not be as familiar with their data as when they manually create a model. Users can brush to select data points of interest. These data points would remain colored in all scatterplots, graying out all other data points. The FI comparison view is correspondingly updated to show local FI of the selected data points, providing further insight. Figure~\ref{fig:lfc} shows FI of points inside the green rectangle of Figure~\ref{fig:psm}. This region is in quadrant 4 and has many orange points. These points were correctly classified by \texttt{LGBM\_2} but incorrectly by \texttt{LogisticRegression\_2}. Further examination of the local FI plots suggests that \texttt{LogisticRegression\_2} might have a tendency of making mistakes by under-weighing the installment and term features.

\section{User Study} 
\label{sec:user study}
Our user study served two purposes: evaluate the design of \tool{}, and use it as a technology probe to understand how AutoML users engage in model comparison tasks. We designed a scenario-based contextual inquiry in which we asked participants to perform the role of a data scientist building an ML model to help loan officers evaluate loan applicants. Participants were presented with 16 models generated by \company's AutoML system (4 optimization variants applied to 4 algorithms). They were asked to use \tool{} to select the best model. To emphasize different comparison criteria, we gave multiple scenarios in which a stakeholder expressed different preferences and asked participants to reconsider their choice. The scenarios included: 1) an executive pointed out that they would prefer an interpretable model; 2) a loan officer commented that it is important that the model's rationale aligns with what features it pays attention to; and 3) an executive emphasized importance of avoiding making loans to people who are likely to default (lower the false positive rate). We asked participants to think aloud as they conducted the comparison tasks and inquired about their thinking when appropriate.

We recruited 14 data scientists from different divisions within an international information technology company. 57.1\% of them reported having 1-5 years of experience working as data scientists (28.6\% over 5 years, 14.3\% less than a year). All were experienced with visualization, and all but one had experience with explainability techniques. 

To familiarize participants with \tool{}, we sent them a tutorial video before the interview. We began the study by asking participants to explore the interface for 5 minutes, then trained them further through a task of identifying the model with the highest accuracy and understanding how another model compares to it. We then gave them the three comparison scenarios in order. After finishing the scenarios, we interviewed participants for 10-15 minutes about their experience with \tool{} and their thoughts on it. Lastly, participants filled out a short survey of our tool, which included the System Usability Scale~\cite{bangor2008empirical}. All interviews were conducted remotely over video conferencing and were recorded and automatically transcribed. We conducted qualitative coding on the interview transcripts while watching the videos to understand users' activities.


\section{Results}
Participants gave high ratings to usability of \tool{}, $M \left(SD\right) = 3.98$ of 5 on the SUS. They found global FI feature to be the most useful, $M \left(SD\right) = 4.29 \left(0.47\right)$, followed by scatterplot matrix, $M \left(SD\right) = 4.21 \left(0.43\right)$, and then the feature to select data points to see local FI, $M \left(SD\right) = 3.93 \left(0.62\right)$. Below, we first describe how participants used the features of \tool{}, then discuss emergent themes on unique design requirements for AutoML model comparison.

\subsubsection{Feature Importance Comparison View}
All participants investigated the global FI and used it to support their choices. For the second scenario to pick a model that is agreeable for loan officers, most participants narrowed down to a small subset of models, and then used global FI to break the tie by favoring models that heavily weighed features such as FICO score or number of trades in the past 2 years. Two participants used FI to verify that the model they intended to select did not exclude important features for the lending domain. Our current visual design compares FI values from all models for one feature. Participants suggested ways to make cross-feature comparison more intuitive, such as by highlighting a selected model's FI values across all feature panels or allowing feature panels to be sorted by FI of a selected model. In some variants, AutoML applies feature transformation and indicates transformation in the name, such as \textit{log\_} or \textit{\_pca\_}. This convention caused some confusion and we noticed that some participants misunderstood situations when a model weighed the transformed feature. It may have been necessary to group related transformed features and provide more information on what AutoML did during optimization. Only a small number of participants explored local FI information. One participant brushed to select data points on which the best candidate made wrong predictions in order to examine why. Some commented that they ``\textit{didn't feel the complexity of the task is high enough to use this},'' since they might not have cared \textit{why} a model made wrong predictions.

\subsubsection{Probability Scatterplot Matrix} Participants welcomed the idea of having all models compared in one visual display and being able to slice the data by brushing. Most participants quickly grasped that the scatterplots could help them compare confidences of models and different types of errors amongst them. Some used it to verify that the model they intended to choose was more confident in its predictions by examining distribution of dots. For scenario of lowering false positives, participants used scatterplots to reason about different types of errors. One participant paid attention to the diagonal of the last-column plots (where the $x$ and $y$ axes are for the same model) and looked for a ``\textit{clean upper right corner}.'' However, the coordinate system was initially confusing for some participants, akin to the finding in~\citet{zhang_manifold_2019} that training is needed for users to understand these plots and what the dot distribution patterns mean. Interpretation of plots is more challenging in the AutoML context, since different ML algorithms with distinct distributions are being compared (some decision tree and random forest models have discrete probabilities, while others have continuous probabilities). Models with discrete probabilities also created visual clutter on the plots. Several participants suggested to show the number of points in each quadrant or brushing selection. Some wished to see raw data when selecting individual points on the plots, or select instances from a data table and highlight them in the plots. These comments suggest that AutoML users are interested in zooming in on specific instances. Since they may not be as familiar with the data when using AutoML compared to hand-crafting a model, a visual display of the instance space could help them identify instances of interest.  
 
\subsubsection{Design Recommendations}
From our qualitative analysis, we identify four areas of user needs around model comparison that should be supported specifically for AutoML.

\textbf{Enable multi-criteria comparison with multiple levels of model details.} Our study demonstrates that the optimal choice in an AutoML search space could be determined by many criteria, which challenges the current practice of AutoML recommending the ``best'' model based solely on performance metrics. While our study intentionally introduced criteria regarding types of errors, interpretability, and model reasoning aligning with domain knowledge, participants incorporated additional criteria such as confidence and reasons for errors. Two participants also commented that computational budget may also be an important criterion. We asked participants if they had experience with model comparison in their own work (not limited to AutoML), and majority confirmed so and commented that it is often done by examining performance only because turn-around time for programmatically-manipulating the data and generating comparative plots is steep. Given the importance of model comparison tasks in AutoML, it is necessary to provide various comparative measures in an interactive and speedy manner. When given a large number of models generated by AutoML, participants used a variety of comparative reasoning strategies: narrowing-down choices, breaking a close tie, reasoning about trade-offs amongst criteria, and verifying a choice to strengthen confidence.

\textbf{Support understanding data.} One user need repeatedly mentioned in interviews is to better understand the data: types of features, their range and distribution, and example data instances. While a lack of knowledge about data is an artifact of the study setup, it could also represent the reality for AutoML users, as they are no longer required to spend time understanding the data. Data scientists frequently utilize model transparency features as lenses to understand their data~\cite{JaimieDrozdal2020,hohman_gamut_2019}. In AutoML, tools such as \tool{} could be the primary place for users to get in touch with their data and retain a sense of agency in the modeling task. 

\textbf{Enable comparison across algorithms and optimization methods.} The set of models generated by AutoML have an innate hierarchical structure as AutoML iteratively applies optimization variants to different ML algorithms. Participants had a tendency of focusing on comparing one selected variation across different algorithms, but also showed interest in understanding how an optimization variant changed a model's behavior. Visualizations that group models by their base algorithm and by optimization variant could allow users to understand the impact of optimization at a glance. However, making comparisons between models that use distinct ML algorithms could impose nuanced design requirements that an AutoML tool should carefully consider in order to be generalizable.

\textbf{Combine algorithmic and procedural transparency.} Participants showed intertwined needs between understanding how AutoML generated models work, and how they were generated, such as which optimization methods were applied and which parameter values were used. This procedural information is necessary for users to make sense of hierarchical structure of models to perform comparative analysis like understanding the meaning of a transformed feature and different rationales between model variants.  When seeing a sub-optimal FI value of a preferred model, a participant expressed interest in ``\textit{tweaking how AutoML does feature engineering},'' suggesting a holistic understanding of algorithmic and procedural operations could facilitate better user control of AutoML.

\section{Conclusion}
\label{sec:discussion}
To support model comparison for AutoML in which users may apply diverse context-specific criteria, \tool{} utilizes linked visualizations of data instances and feature importance. Our study highlights the need to help AutoML users understand detailed behaviors of machine-created models, and shows users' complex reasoning strategies and nuanced requirements resulting from the unique structure and building process of AutoML models. Future work should explore supporting a more structured model comparison workflow to help users navigate these complexities.  

\begin{acks}
This work was supported by the Rensselaer-IBM AI Research Collaboration (http://airc.rpi.edu), part of the IBM AI Horizons Network (http://ibm.biz/AIHorizons).
\end{acks}

\bibliographystyle{ACM-Reference-Format}
\bibliography{acmart}

\end{document}